# Soft materials for linear electromechanical energy conversion

Antal Jákli[1] and Nándor Éber[2]

[1]Liquid Crystal Institute and Chemical Physics Interdisciplinary Program,
Kent State University, Kent, OH 44242, USA

[2]Institute for Solid State Physics and Optics, Wigner Research Centre for Physics,
Hungarian Academy of Sciences, H-1525 Budapest, P.O. Box. 49, Hungary

**Abstract**:

*We briefly review the literature of linear electromechanical effects of soft materials, especially in synthetic and biological polymers and liquid crystals (LCs). First we describe results on direct and converse piezoelectricity, and then we discuss a linear coupling between bending and electric polarization, which maybe called bending piezoelectricity, or flexoelectricity.*

## I. Piezoelectricity

Piezoelectricity is a linear coupling between electric and mechanical properties in special non-centrosymmetric materials. It has been known in crystals since the work of Pierre and Jacques Curie in 1880[1]. Piezoelectric properties were also found in amorphous and polycrystalline materials, such as ferroelectric ceramics.[2,3,4] The existence of piezoelectricity in certain synthetic and biological polymers[5], such as bone and tendon has been extensively studied.[6,7] Piezoelectric effects in polymers are generally small, but can be increased when they are subjected to a strong DC electric field at elevated temperatures.[8] Due to their flexibility and the possibility to prepare films of large area, these materials (especially polyvynilidene fluoride, PVDF) have been utilized as the active element in many applications ranging from infrared detectors to loudspeakers. Certain cellular polymers internally charged by corona discharge (ferroelectrets) have been found to behave like soft sensitive piezoelectrics.[9,10,11] Similarly large and soft piezoelectricity was found in fiber mats composed of ferroelectric barium titanate (BT) nanoparticles dispersed in polylactic acid (PLA).[12]

As the type of materials with experimentally observed piezoelectric coupling, so the definition of piezoelectricity extended considerably in time. Originally it was used only for crystals in connection with compressions, but later it was generalized to polymers and other materials for any strains and stresses, including shear. The *direct piezoelectric effect* (when electric polarization is produced by mechanical stress) mathematically can be expressed by the equation:

$$P_i = \sum_{jk} d_{i,jk} T_{jk},  \quad (1)$$

where $P_i$ is the *i*th component of the polarization induced by the stress tensor $T_{jk}$. The third rank tensor coefficients $d_{i,jk}$ are called piezoelectric coefficients.

In the *converse* (or "inverse") *effect* the material becomes strained when electric field is applied. Mathematically this is expressed as

$$S_{jk} = \sum_i d_{i,jk} E_i. \quad (2)$$

Here $S_{jk} = \partial s_j / \partial x_k$ is the strain tensor ($s_j$ is the displacement of a volume element from its equilibrium position), and $E_i$ is the external electric field.

By symmetry not only solids, but also liquid crystals (LC) can lack inversion symmetry due to the chirality of rod-shape molecules containing carbon stereo-centers, or due to asymmetric packing of molecules with special (bent, banana) shapes.[13] The chiral liquid crystal phases are the cholesteric (N*), chiral smectic (e.g. SmC* of rod-shape molecules,[14] or SmCP of bent-core molecules[15]) and the chiral columnar phases.

| Phase | Molecular shape | Symmetry | Non-zero Piezo constants |
|---|---|---|---|
| N*, SmA* | Chiral rod | $D_\infty$ | $d_{1,23} = -d_{2,13}$ |
| SmAP | Bent shape | $C_{2v}$ | $d_{3,11}; d_{3,22}; d_{3,33}, d_{1,13}; d_{2,23}$ |
| SmC* | Chiral rod | $C_2$ | $d_{3,11}; d_{3,22}; d_{3,33}, d_{1,13}; d_{2,23}; d_{1,23}; d_{2,13}; d_{3,12}$ |
| SmCP | Bent shape | $C_2$ | $d_{3,11}; d_{3,22}; d_{3,33}, d_{1,13}; d_{2,23}; d_{1,23}; d_{2,13}; d_{3,12}$ |
| SmCG | Bent shape | $C_1$ | All 18 $d_{i,jk}$ |
| Col*$_{tilt}$ | Chiral disc | $C_2$ | $d_{3,11}; d_{3,22}; d_{3,33}, d_{1,13}; d_{2,23}; d_{1,23}; d_{2,13}; d_{3,12}$ |
| P$_h$ | Bowl shape | $C_{\infty v}$ | $d_{3,33}; d_{3,11}=d_{3,22}$ |
| P*$_h$ | Chiral bowl | $C_\infty$ | $d_{3,33}; d_{3,11}; d_{3,22}$ |
| P$_{tilt}$ | Tilted bowl | $C_2$ | $d_{3,11}; d_{3,22}; d_{3,33}, d_{1,13}; d_{2,23}; d_{1,23}; d_{2,13}; d_{3,12}$ |

*Table 1: List of the most important liquid crystalline phases with their symmetry and non-vanishing piezoelectric coupling constants.*

*Table 1* lists the LC phases that have no inversion symmetry, i.e., which can have linear couplings between electric field and mechanical strain. We see that several of them (SmC* of chiral rod-shape, the tilted columnar phase of chiral disc shape and the SmCP of achiral bent-core or tilted bowl-shape molecules) have $C_2$ symmetry, which allows 8 independent piezoelectric coefficients.

In the SmA* and N* phases a shear induces a tilt, which, combined with the lack of mirror plane due to the molecular chirality, allows the existence of a polar vector normal to the shear plane. This is illustrated in the upper part of Figure 1.

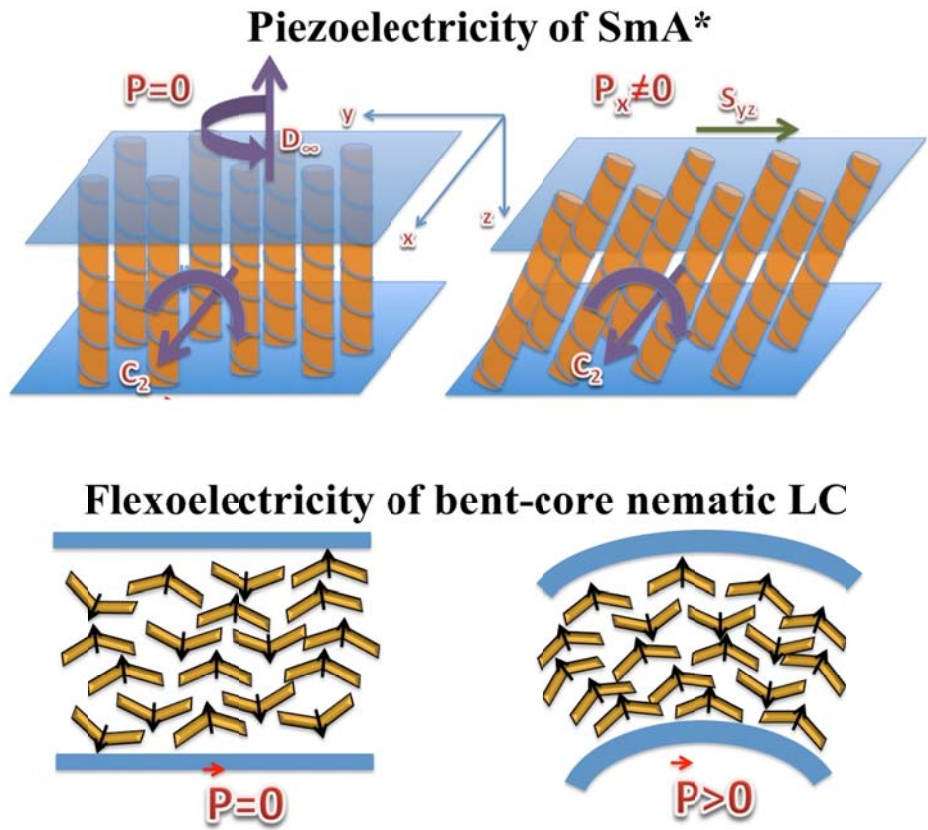

*Figure 1: Illustration of the main physical mechanism of piezoelectricity and flexoelectricity in LCs.*

The inverse effect, when an electric field is applied perpendicular to the director, results in a shear strain, which would also lead to a tilt of the director. The optical consequence of this effect, rotation of the optic axis (the director) proportional to the electric field, is known as electro-clinic effect.[16,17,18]

The **_direct piezoelectric_** (or linear mechano-electrical) effect in SmC* materials is due to the shear-induced polarization[19,20], and is arising from the distortion of the helix. In the ground state (no-shear) the polarization averages out in bulk, but shear unwinds the helix and leads to a polarization normal to the shear plane. Direct piezoelectric effects were also observed in various biological membranes[21] and in aqueous lyotropics, where the highly conductive water is often replaced by ethylene glycol.[22,23,24] Linear mechano-electrical effects were also found on non-aqueous lyotropic phases[21,25] Recently neat phospholipids were also shown to be piezoelectric.[26,27]

The **_converse piezoelectric_** (linear electro mechanical) effect in SmC* materials[28,29,30] may result in audible acoustic effects implying their possible use in electromechanical transducers.[31,32] A systematic study[33] with accurate control of the alignment revealed that the vibrations parallel to the smectic layers and the film surface are generally the strongest, especially if the polarization is also parallel to the plates.

Linear electro-mechanical effects of free-standing ferroelectric SmC* films were studied by light scattering while applying AC electric fields along the film surface.[34,35,36,37] In addition, they were also investigated in ferroelectric columnar LCs,[38] SmC* elastomers[39,40] and glasses,[41] in various biological systems,[42] lyotropic LCs and in membranes.[21] Converse piezoelectricity was also invoked[43] to explain the swelling of membranes in response to voltage changes.[44] Converse piezoelectric measurements in ferroelectric SmCP bent-core LCs[45], and a summary of experimental studies in SmC* elastomers[46] have been published only recently. An electro-clinic effect related field-induced twist, whose direction changes sign with that of the electric field, was also reported on SmA* elastomers.[47,48]

**II. Flexoelectricity**

Another linear electromechanical phenomenon, flexoelectricity, connects bending and electricity. Initially it was regarded as bending piezoelectricity[49,50], however, in contrast to piezoelectricity, flexoelectricity does not require the lack of inversion symmetry and the coupling constants $e_{ijkl}$ are given by a fourth rank tensor:

$$P_i^{flexo} = e_{ijkl}\frac{\partial S_{jk}}{\partial x_l}; \quad A\frac{\partial T_{ij}}{\partial x_k} = e_{ijkl}E_l, \quad (3)$$

corresponding to direct and converse flexoelectricity, respectively. $S_{jk}$ ($T_{ij}$) are the elements of the strain (stress) tensor, and $A$ is the area of the sample. In crystals the first experiments [51] in 1988 gave e ~ $10^{-11}$–$10^{-10}$ C/m. In perovskite-type [52] and relaxor ferroelectric ceramics [53,54,55] and in polyvinylidene fluoride (PVDF) film [56,57] the flexoelectric coefficients are 4 orders of magnitude larger than in dielectric crystals, and the large polarization induced by flexing is evidently of ferroelectric origin.

In fluid LCs stresses and strains are related mostly to orientational elasticity. R.B. Meyer predicted Flexoelectricity[50] for nematic LCs composed of dipolar molecules with asymmetric shapes, such as pear-shape or bent-core. In the absence of electric fields the molecular dipoles average out and the net polarization is zero. However, when the director is subjected to splay or bend deformations, respectively, the material become macroscopically polar:

$$P^{flexo} = e_1 \mathbf{n} (\text{div } \mathbf{n}) - e_2 \mathbf{n} \times (\text{curl } \mathbf{n}) \quad (4)$$

This effect is illustrated in the lower row of Figure 1.

Molecular statistical theories based on molecular packing considerations[58,59,60] predicted for the splay and bend flexo-coefficients $e_1$ and $e_3$ to be about 1 pC/m for nearly rod-shape molecules, and about an order of magnitude larger $e_3$ values for banana shape molecules. Actual measurements on rod-shape LCs [61,62,63,64,65,66,67,68,69] typically give about $e_1$~$e_3$~10pC/m, due to a contribution of a quadrupolar mechanism.[70] These values are similar to those measured on elastomers subjected to a deformation gradient.[51]

A "peculiar kind of piezoelectric effect" was measured on black lipid membranes (BLM)[71,72,73] and found that the molecular basis of the flexoelectricity of lipid bilayers is an asymmetric redistribution of charges, dipoles and the splayed uniaxial orientation of their quadrupolar moments. [74] Converse flexoelectric measurements on BLM [75] and vesicles[76] found $e_1$ ~ 100 pC/m, which is an order of magnitude larger than typical for calamitic thermotropic LCs. The sensory mechanism of outer hair cell composite membranes[77,78] can be understood by the flexoelectric properties of the lipid bilayer.[79] The converse of this effect, i.e., voltage-generated curvature has also been observed.[80]

The bend flexoelectric coefficient determined from mechanical bending induced electric current measurements in bent-core LCs[81,82] show 3 orders of magnitudes larger values than in normal dielectric LCs. An example on 4-chloro-1,3-phenylene bis 4-[4'-(9-decenyloxy) benzoyloxy] benzoate (ClPbis10BB) in comparison with a typical rod-shape liquid crystal pentyl cyano biphenyl (5CB) is shown in Figure 2.

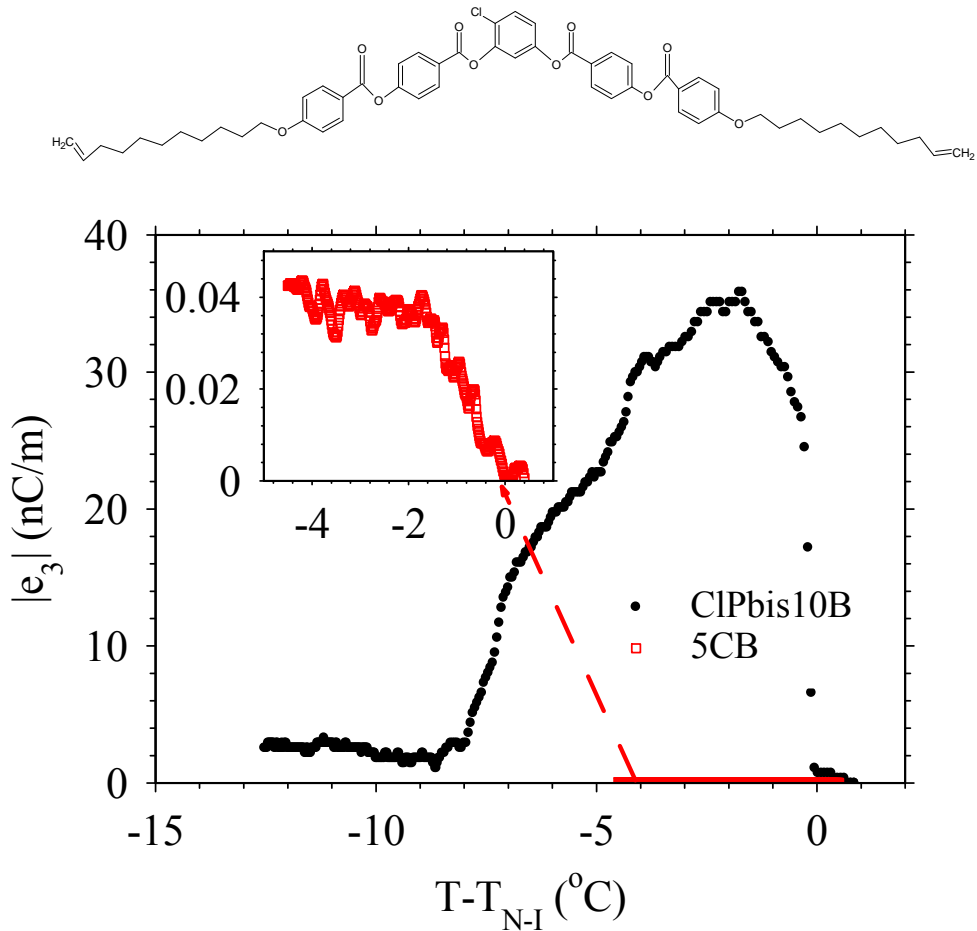

Figure 2: *Variation of the flexoelectric coefficient on a relative temperature scale $T-T_{N-I}$ for the bent-core liquid crystal ClPbis10BB (molecular structure seen above the graph), and for the calamitic liquid crystal 5CB measured in cells of $A=1cm^2$ active areas. The inset shows part of the figure ($|e_3|$ of 5CB) at a magnified scale.*

This increase is similar to that of ferroelectric ceramics[55] and indicates the presence of ferroelectric smectic clusters[83,84,85,86] in bent-core nematics.

**Conclusion:**

The physical effects reviewed above represent couplings between electrical and mechanical phenomena, which might allow energy conversion.

These concepts turned out to be useful in interpreting some functions of living organisms, successfully developed during evolution. These effects are already utilized in state-of-the-art sensor and actuator technologies, and might have potential for further practical applications. The latter hopefully prompt designing and synthesizing more appropriate functional materials (presumably polymers or elastomers) with larger coupling coefficients and easier processability.

### III. References

[1] J. Curie and P. Curie, Dévelopment par compression de l'électricité polaire dans les cristaux hémlédres a fáces inclinees. *Bulletin no.4 de la Societee Mineralogique de France*, **3**, 90 (1880)

[2] B.M. Wul and I.M. Goldman, Dielectric Constant of $BaTiO_3$ as a Function of Strength of an Alternating Field, *Dokl. Akad. Nauk. SSSR*, **49**, 179 (1945)

[3] A. Von Hippel, R.G. Breckenridge, F.G. Chesley and L. Tisza, High dielectric constant ceramics, *Ind. Eng. Chem.*, **38**, 1097-1109 (1946)

[4] S. Roberts, Dielectric and Piezoelectric Properties of Barium Titanate *Phys. Rev.*, **71**, 890-895 (1947)

[5] Y. Wada, Chapter 4. Piezoelectricity and Pyroelectricity, In *Electronic properties of polymers* ed. J. Most and G. Pfister, Wiley-Interscience, New York (1982)

[6] E. Fukada, Piezoelectricity in polymers and biological materials. *Ultrasonics*, **6**, 229 (1968)

[7] E. Fukada, Piezoelectric properties of biological macromolecules. *Adv. Biophys.*, **6**, 121 (1974)

[8] H. Kawai, The piezoelectricity of poly(vinylidene fluoride). *Jpn. J. Appl. Phys.*, **8**, 975-976 (1970)

[9] J. Lekkala, R. Poramo, K. Nyholm and T. Kaikkonen, EMF force sensor–a flexible electret film for physiological applications, *Med. Biol. Eng. Comput.*, **34**, 67-68 (1996)

[10] G. M. Sessler and J. Hillenbrand, Electromechanical response of cellular electret films, *Appl. Phys. Lett.*, **75**, 3405-3407 (1999)

[11] S. Bauer, R. Gerhard-Multhaupt and G. M. Sessler, Ferroelectrets: Soft Electroactive Foams for Transducers, *Physics Today,* **57**, 37 (2004)

[12] (*) J. Morvan, E. Buyuktanir, J.L. West and A. Jákli, Highly piezoelectric biocompatible and soft composite fibers, *Appl. Phys. Lett.*, **100**, 063901/1-4 (2012) *Biocompatible soft fibers were prepared from poly lactic acid containing barium titanate ferroelectric ceramic particles by electrospinning. The fibers showed two orders of magnitude larger piezoelectric constant per weight than single crystal barium titanate films.*

[13] T. Niori, T. Sekine, J. Watanabe, T. Furukawa and H. Takezoe, Distinct ferroelectric smectic liquid crystals consisting of banana shaped achiral molecules, *J. Mater. Chem.*, **6(7)**, 1231-1233, (1996)

[14] R.B. Meyer, L. Liebert, I. Strzelecki and P. Keller, Ferroelectric liquid crystals, *J. Phys. Lett. (Paris)*, **36**, L69-L72 (1975)

[15] D.R. Link, G. Natale, R. Shao, J.E. Maclennan, N.A. Clark, E. Körblova and D.M. Walba, Spontaneous formation of macroscopic chiral domains in a fluid smectic phase of achiral molecules, S*cience,* **278**, 1924-1927 (1997)

[16] S. Garoff and R.B. Meyer, Electroclinic Effect at the *A-C* Phase Change in a Chiral Smectic Liquid Crystal, *Phys. Rev. Lett.*, **38**, 848-851 (1977)

[17] C.H. Bahr and G. Heppke, Optical and dielectric investigations on the electroclinic effect exhibited by a ferroelectric liquid crystal with high spontaneous polarization, *Liquid Crystals*, **2**, 825-831 (1987)

[18] G. Andersson, I. Dahl, P. Keller, W. Kuczynski, S.T. Lagerwall, K. Skarp and B. Stebler, Submicrosecond electro-optic switching in the liquid-crystal smectic *A* phase: The soft-mode ferroelectric effect, *Appl. Phys. Lett.*, **51** (9), 640-642 (1987)

[19] P. Pieranski, E. Guyon and P. Keller, Shear flow induced polarization in ferroelectric smectics C, *J. Phys. (Paris)*, **36**, 1005-1010 (1975)

[20] A. Jákli and L. Bata, Mechano-Electrical Effects on Planar $S_C$* Liquid Crystals, *Mol. Cryst. Liq. Cryst.*, **201**, 115-124 (1991)

[21] L.M. Blinov, S.A. Davidyan, A.G. Petrov and S.V. Yablonsky, Manifestation of ferroelectricity in a lyotropic liquid crystal with a chiral impurity: a structural analog of a biological membrane, *JETP Lett.*, **48**, 285-288 (1988)

[22] N. Moucharafieh and S. Friberg, *Mol.Cryst.Liq.Cryst.*, **49**, 231 (1979)

[23] A.G. Petrov and G. Durand, Thermal instability in lamellar phases of lecithin: a planar undulation model, *J. Physique Lett.*, **44**, L-793-L-798 (1983).

[24] A.G. Petrov, M. Cagnon, Y. Galerne and G. Durand, Thermal and Mechanical Instabilities in Nonaqueous Lamellar Lyotropic Lecithin-Ethylenglycol, *Mol. Cryst. Liq.Cryst.*, **154**, 179-193 (1988).

[25] Y. Kagawa and T. Hatakeyama, Piezoelectric effect in liquid crystals, *J. Sound Vibration*, **53**, 585-593 (1977)

[26] A. Jákli, J. Harden, C. Notz and C. Bailey, Piezoelectricity of phospholipids: a possible mechanism for mechanoreception and magnetoreception in biology, *Liq. Cryst.*, **35** (4), 395-400 (2008)

[27] J. Harden, N. Diorio, A.G. Petrov, A. Jákli, Chirality of lipids makes fluid lamellar phases piezoelectric, *Phys. Rev. E*, **79**, 011701 (2009).

[28] A. Jákli, L. Bata, Á. Buka, N. Éber and I. Jánossy, New electromechanical effect in chiral smectic C* liquid crystals, *J. Physique Lett.*, **46**, L-759-L-761 (1985)

[29] A. Jákli, L. Bata, Á. Buka, N. Éber, Electromechanical effect in $S_C$* liquid crystals. *Ferroelectrics*, **69**, 153-163 (1986)

[30] A.P Fedoryako, M.N. Pivnenko, E.V. Popova, V.P. Seminozhenko, Free oscillations in a ferroelectric liquid crystal, TECHNICAL PHYSICS LETTERS, **23**, 287-288 (1997)

[31] L. Bata, N. Éber and A. Jákli, Electromechanical transducer, *Hungarian Patent,* **20052B** (10. 28. 1988)

[32] K. Yuasa and K. Hashimoto, (Idemitsu Kosan Co., Ltd.) *Jpn. Kokai Tokyo Koho* JP **01,175,400** (11 Jul 1989); Seiko Epson (**62-203132**)

[33] A. Jákli and A. Saupe, The Role of Goldstone Mode and Electroclinic Effects in Electromechanical Responses of Chiral Smectic C Liquid Crystals, *Mol. Cryst. Liq. Cryst.*, **237**, 389-398 (1993)

[34] E.I. Demikhov, S.A. Pikin and E.S. Pikina, Kink switching in ferroelectric free-standing films with high spontaneous polarization, *Phys. Rev. E,* **52**, 6250-6259 (1995)

[35] S. Uto, M. Ozaki and K. Yoshino, Molecular reorientation and deformation of a freely suspended ferroelectric liquid crystal film, *Appl. Phys. Lett.,* **74**, 117-119 (1999)

[36] S. Uto, E. Tazoh, M. Ozaki and K. Yoshino, Mechanical vibration of freely suspended ferroelectric liquid-crystal film excited by sound and electric field, J. Appl. Phys., **82**, 2791-2794 (1997)

[37] S.V. Yablonskii, T. Oue, H. Nambu, A.S. Mikhailov, M. Ozaki and K. Yoshino, Electromechanical effect in freely suspended liquid crystal films, *Appl. Phys. Lett.*, **75** (1), 64-66 (1999)

[38] A. Jákli, M. Müller, D. Krüerke and G. Heppke, First observation of electromechanical effects in a chiral ferroelectric columnar liquid crystal, *Liq. Cryst.,* **24**, 467 (1998)

[39] S.U. Vallerien, F. Kremer, E.W. Fischer, H. Kapitza, R. Zentel and H. Poths, Experimental proof of piezoelectricity in cholesteric and chiral smectic C*-phases of LC-elastomers, *Makromol. Chem. Rapid Commun.,* **11**, 593-598 (1990)

[40] C.-C. Chang, L.-C. Chien and R.B. Meyer, Piezoelectric effects in cholesteric elastomer gels *Phys. Rev. E,* **55**, 534-537 (1997)

[41] A. Jákli, T. Toth-Katona, T. Scharf, M.. Schadt and A. Saupe, Piezoelectricity of a ferroelectric liquid crystal with a glass transition, *Phys. Rev. E*, **66**, 011701 (2002)

[42] W.S. Williams, *Ferroelectrics*, **41**, 2251 (1982) and references therein

[43] H.R. Leuchtag, Indications of the existence of ferroelectric units in excitable-membrane channels, *J. Theor. Biol.*, **127**, 321 (1987)

[44] K. Iwasa, I. Tasaki, R.C. Gibbons, Swelling of nerve fibers associated with action potentials, *Science*, **210**, 338 (1980)

[45] A. Jákli, I.C. Pinte, J.L. Serrano, M.B. Ros and M.R. de la Fuente, Piezoelectric and Electric-Field-Induced Properties of a Ferroelectric Bent-Core Liquid Crystal, *Adv. Mater*., **21**, 3784–3788 (2009)

[46] K. Hiraoka, M. Kobayashi and R. Kazama, Electromechanics of Monodomain Chiral Smectic C Elastomer: Mechanical Response to Electric Stimulation, 5th International Liquid Crystal Elastomer Conference, September 24-26, 2009, Kent, OH, USA, *Electronic Liquid Crystal Presentations*: http://www.e-lc.org/presentations/docs/ 2009_10_02_05_33_52

[47] C.M. Spillmann, B.R. Ratna and J. Naciri, Anisotropic actuation in electroclinic liquid crystal elastomers, *Appl. Phys. Lett.*, **90,** 021911 (2007)

[48] B.R Ratna , J.R Deschamps , J.H Konnert , C.M Spillmann and J. Naciri, Stress and E-field Induced Structural Changes in Electroclinic Elastomers, 5th International Liquid Crystal Elastomer Conference, September 24-26, 2009, Kent, OH, USA; http:/www.e-lc.org/presentations/docs/2009_10_05_13_48_26

[49] Sh.M. Kogan, Piezoelectric effect on non-uniform deformation and acoustical scattering of carriers in crystals, *Sov. Phys. Solid State*, **5** (1964) 2069-2071

[50] R.B. Meyer, Piezoelectric effects in liquid crystals, *Phys. Rev. Lett.*, **22** (1969) 918-921

[51] M. Marvan and A. Havranek, Flexoelectric effect in elastomers, In ed. I. Chudácek, *Relationships of Polymeric Structure and Properties, Progress in Colloid and Polymer Science*. Vol. **78**, Springer-Verlag, New York, 1988. pp. 33–36.

[52] (**) A. Jákli, Electro-mechanical effects in liquid crystals, *Liquid Crystals,* **37** (6), 825-837 (2010)
*Linear (piezo- and flexoelectricity) and quadratic (electrostriction) electromechanical effects of liquid crystals (nematic, smectic and columnar phases) and liquidcrystalline polymers and elastomers are reviewed.*

[53] W. Ma and L. Cross, *Appl. Phys. Lett.*, An experimental investigation of electromechanical response in a dielectric acrylic elastomer, Applied Physics A, **78**, 1201-1204 (2004)

[54] W. Ma and L.E. Cross, A. Large, Large flexoelectric polarization in ceramic lead magnesium niobate, *Appl. Phys. Lett.,* **79**, 4420-4422 (2001).

[55] Y. Ma, Flexoelectricity: strain gradient effects in ferroelectrics, *Physica Scripta*, **T129**, 180-183 (2007)

[56] (*) S. Baskaran, N. Ramachandran, X. He, S. Thiruvannamalai, H. J. Lee, H. Heo, Q. Chen and J.Y. Fu, Giant flexoelectricity in polyvinylidene fluoride films, *Physics Letters A*, **375**, 2082-2084 (2011)
*A polyvinylidene fluoride film was found to exhibit giant flexoelectric polarization upon mechanical deformation.*

[57] (*) S. Baskaran, X. He, Q. Chen and J.Y. Fu, Experimental studies on the direct flexoelectric effect in α-phase polyvinylidene fluoride films, *Appl. Phys. Lett.,* **98**, 242901 (2011).
*Giant flexoelectric response was detected in the apolar α-phase of the polymer film of polyvinylidene fluoride. The strain gradients occurring in the samples were numerically calculated.*

[58] W. Helfrich, A simple method to observe the piezoelectricity of liquid crystals, *Phys. Lett.,* **35A**, 393-394 (1971)

[59] W. Helfrich, The strength of piezoelectricity in liquid crystals, *Z. Naturforsch*, **26a**, 833-835 (1971)

[60] A. Derzhanski and A.G. Petrov, A molecular-statistical approach to the piezoelectric properties of nematic liquid crystals, *Phys. Lett.,* **36A**, 483-484 (1971)

[61] D. Schmidt, M. Schadt and W. Helfrich, Liquid-crystalline curvature electricity: The bending mode of MBBA, *Z. Naturforsch,* **27a**, 277-280 (1972)

[62] G. Barbero, P. Taverna Valabrega, R. Bartolino and B. Valenti, Evidence for the flexo-electric effect in a nematic liquid crystal cell, *Liquid Crystals*, **1**, 483-490 (1986)

[63] I. Dozov, Ph. Martinot-Lagarde and G. Durand, Flexoelectrically controlled twist of texture in a nematic liquid crystal, *J. Phys. Lett. (Paris)*, **43**, L-365-L-369 (1982)

[64] B. Valenti, C. Bertoni, G. Barbero, P. Taverna Valabrega and R. Bartolino, Flexoelectricity in the hybrid aligned nematic cell, *Mol. Cryst. Liq. Cryst.*, **146**, 307-320 (1987)

[65] S. Warrier and N.V. Madhusudana, An AC electrooptic technique for measuring the flexoelectric coefficient ($e_1+e_3$) and anchoring energies of nematics, *J. Phys. II (France)*, **7**, 1789-1803 (1997)

[66] L. M. Blinov, M. I. Barnik, H. Ohoka, M. Ozaki and K. Yoshino, Separate measurements of the flexoelectric and surface polarization in a model nematic liquid crystal p-methoxybenzylidene-p'-butylaniline: Validity of the quadrupolar approach, *Phys. Rev. E*, **64**, 031707 (2001)

[67] N.T. Kirkman, T. Stirner and W.E. Hagston, Continuum modeling of hybrid-aligned nematic liquid crystal cells: Optical response and flexoelectricity-induced voltage shift, *Liquid Crystals*, **30**, 1115-1122 (2003)

[68] S. Ponti, P. Ziherl, C. Ferrero and S. Zumer, Flexoelectro-optic effect in a hybrid nematic liquid crystal cell, *Liquid Crystals*, **26**, 1171-1177 (1999)

[69] T. Takahashi, S. Hashidate, H. Nishijou, M. Usui, M. Kimura and T. Akahane, Novel measurement method for flexoelectric coefficients of nematic liquid crystals, *Jpn. J. Appl. Phys.*, **37**, 1865-1869 (1998)

[70] J.P. Marcerou and J. Prost, On the microscopic interpretation of flexoelectricity, *J. Phys. France,* **38**, 315-324 (1977)

[71] V.I. Passechnik and V.S. Sokolov, Permeability change of modified bimolecular phospholipid membranes accompanying periodical expansion, *Biofizika (Moscow)*, **18**, 655-660 (1973)

[72] A.G. Petrov and A. Derzhanski, Membranes and lyotropic phases on some problems in the theory of elastic and flexoelectric effects in bilayer lipid membranes and biomembranes, *J. Phys. Colloques* **37**, C3-155-160 (1976)

[73] A.G. Petrov, Flexoelectricity of lyotropics and biomembranes, *Il Nuovo Cimento*, **3D**, 174-192 (1984)

[74] A.G. Petrov and V.S. Sokolov, Curvature-electric effect in black lipid membranes, *Eur. Biophys. J.*, **133**, 139-155 (1986)

[75] A. Derzhanski, A.G. Petrov, A.T. Todorov and K. Hristova, Flexoelectricity of lipid bilayers, *Liquid Crystals,* **7**(3), 439-449 (1990)

[76] L-T. Gao, X-Q. Feng, Y-J. Yin and H. Gao, An electromechanical liquid crystal model of vesicles, *J. Mech. Phys. Solids*, **56**, 2844-2862 (2008)

[77] R.M. Raphael, A.S. Popel and W.E. Brownell. A membrane bending model of outer hair cell electromotility, *Biophys. J.,* **78**, 2844-2862 (2000)

[78] A.A. Spector, N. Deo, K. Grosh, J.T. Ratnanather and R.M. Raphael, Electromechanical models of the outer hair cell composite membrane, *J. Membrane Biol.,* **209**, 135-152 (2006)

[79] A.G. Petrov, Flexoelectric model for active transport, in *Physical and chemical bases of biological information transfer, (J. Vassileva, ed), Plenum Press, New York, pp. 167 (1975)*

[80] A.T. Todorov, A.G. Petrov and J.H. Fendler, First observation of the converse flexoelectric effect in bilayer lipid membranes*, J. Phys. Chem.,* **98**, 3077-3099 (1994)

[81] J. Harden, B. Mbanga, N. Éber, K. Fodor-Csorba, S. Sprunt, J. T. Gleeson and A. Jákli, Giant flexoelectricity of bent-core nematic liquid crystals, *Phys. Rev. Lett.*, **97**, (2006), 157802

[82] M. Chambers, R. Verduzco, J.T. Gleeson, S. Sprunt and A. Jákli, Flexoelectricity of a calamitic liquid crystal elastomer swollen with a bent-core liquid crystal, *J. Mater. Chem.*, **19** (42), 7909-7913 (2009)

[83] C. Bailey, K. Fodor-Csorba, J.T. Gleeson, S.N. Sprunt and A. Jakli, Rheological properties of bent-core liquid crystals, *Soft Matter,* **5**, 3618 (2009).

[84] N. Vaupotic, J. Szydlowska, M. Salamonczyk, A. Kovarova, J. Svoboda, M. Osipov, D. Pociecha and E. Gorecka, Structure studies of the nematic phase formed by bent-core molecules, *Phys. Rev. E,* **80**, 030701 (2009)

[85] O. Francesangeli, V. Stanic, S. I. Torgova, A. Strigazzi, N. Scaramuzza, C. Ferrero, I. P. Dolbnya, T. M. Weiss, R. Berardi, L. Muccioli, S. Orlandi and C. Zannoni, Ferroelectric Response and Induced Biaxiality in the Nematic Phase of Bent-Core Mesogens, *Adv. Funct. Mater,* **19**, 1 (2009).

[86] (*) P. Salamon, N. Éber, Á. Buka, J. T. Gleeson, S. Sprunt and A. Jákli, Dielectric properties of mixtures of a bent-core and a calamitic liquid crystal, *Phys. Rev. E*, **81**, 031711 (2010)

*Dielectric spectroscopy of a bent-core nematic liquid crystal proved the existence of a dielectric relaxation at unusually low frequencies. The dielectric spectra of mixtures of bent-core and calamitic nematics turned out to be superpositions of the spectra of the individual components.*